# Expanders via Local Edge Flips

Zeyuan Allen-Zhu[*1], Aditya Bhaskara[†2], Silvio Lattanzi[‡2], Vahab Mirrokni[§2], and Lorenzo Orecchia[¶3]

[1]Princeton University
[2]Google Research NYC
[3]Boston University

October 26, 2015

## Abstract

Designing distributed and scalable algorithms to improve network connectivity is a central topic in peer-to-peer networks. In this paper we focus on the following well-known problem: given an $n$-node $d$-regular network for $d = \Omega(\log n)$, we want to design a decentralized, local algorithm that transforms the graph into one that has good connectivity properties (low diameter, expansion, etc.) without affecting the sparsity of the graph. To this end, Mahlmann and Schindelhauer introduced the random "flip" transformation, where in each time step, a random pair of vertices that have an edge decide to 'swap a neighbor'. They conjectured that performing $O(nd)$ such flips at random would convert any connected $d$-regular graph into a $d$-regular expander graph, with high probability. However, the best known upper bound for the number of steps is roughly $O(n^{17}d^{23})$, obtained via a delicate Markov chain comparison argument.

Our main result is to prove that a natural instantiation of the random flip produces an expander in at most $O(n^2 d^2 \sqrt{\log n})$ steps, with high probability. Our argument uses a potential-function analysis based on the matrix exponential, together with the recent beautiful results on the higher-order Cheeger inequality of graphs. We also show that our technique can be used to analyze another well-studied random process known as the 'random switch', and show that it produces an expander in $O(nd)$ steps with high probability.

---

[*]Email: `zeyuan@csail.mit.edu`.

[†]Email: `bhaskaraaditya@gmail.com`.

[‡]Email: `silviol@google.com`.

[§]Email: `mirrokni@google.com`.

[¶]Email: `orecchia@bu.edu`.

# 1  Introduction

Graph processes that are *local* (i.e., involve communication only between neighbors) have received a lot of attention in the past few years due to their important applications in distributed systems. Thanks to this effort we now have a better understanding of some fundamental problems like gossiping and load balancing [12, 13, 16, 31]. Nevertheless, there are still fundamental questions about which our knowledge is limited. In this paper we focus on one of these questions: can we design a local graph transformation that, if applied multiple times to an input undirected graph $G$, can *efficiently* transform the graph into a *well-connected* one without increasing the number of edges?

Well-connectedness is a particularly important requirement for dynamic networks such as sensor networks and peer-to-peer (P2P) networks, where it is fundamental to maintain a network with a small diameter and a robust structure against external attacks. For those networks, it has been shown [18, 35] that regular random graphs are excellent from various prospectives. The key property of regular random graphs is *expansion*, which implies a number of other desirable properties, e.g. logarithmic diameter, high vertex connectivity and a small mixing time of random walks [22]. In this paper, we thus study the question of obtaining an *expander graph* efficiently, using only local changes to the graph.

The problem of generating well-connected graphs using *simple* (involving only a small set of vertices and edges between them) transformations is not new: a first and nice method was introduced by McKay [24] with the so-called *simple-switch* transformation. In this transformation one selects a random pair of edges $(r, i)$ and $(s, j)$ of the graph $G$ and, if the edges $(r, j)$ and $(s, i)$ do not exist in the graph and $r \neq j$ and $s \neq i$, then the edges $(r, i)$ and $(s, j)$ are deleted from the graph and the edges $(r, j)$ and $(s, i)$ are added. Interestingly, it is possible to show that by repeatedly applying this transformation on $d$-regular graph, it converges to a random $d$-regular graph and furthermore the convergence time of this process has been extensively studied [8, 25]. Recently Greenhill [15] proved that this "switch Markov chain" (the Markov chain on the set of all $d$-regular graphs which is induced by the switch operation) is also rapidly mixing for non-regular graphs.

The simple-switch transformation is very elegant, but it has two fundamental limitations: (1) in practice, in a large distributed system, sampling a random pair of edges is a challenging (and highly non-local) task, and (2) there is a small probability that the simple switch transforms a connected graph into a disconnected one (this issue is also crucial in P2P and sensor networks applications).

For this reason Mahlmann and Schindelhauer introduced [22] a *local* variation of the simple switch, known as the *flip* transformation. In this transformation, one selects a random length three path on four nodes $i, r, s, j$, such that (1) $(i, r), (r, s)$ and $(s, j)$ are edges of $G$, (2) edges $(r, j)$ and $\{s, i\}$ do not exist in the graph, and (3) $r \neq j$ and $s \neq i$. Then, one deletes the edges $(r, i)$ and $(s, j)$ from the graph and adds the edges $(r, j)$ and $(s, i)$. Interestingly, Mahlmann and Schindelhauer showed that, starting from any graph $G$, by repeatedly applying the flip transformation it is possible to obtain a random $d$-regular graph. Later, Feder *et al.* [11] showed that the mixing time of the "flip Markov chain" is polynomially bounded. More recently, Cooper, Dyer, and Handley [7] showed a better upper bound: they proved that the number of flip transformations needed is $O(n^{16}d^{22}(dn \log dn + \log \varepsilon^{-1}))$ to get $\varepsilon$ close to a random $d$-regular graph (and therefore an expander). The analyses in [7] and [11] are based on the idea of simulating a simple-switch transformation using a sequence of flips, and then analyzing the resulting flip Markov chain.

In a sharp contrast to the proven upper bounds, it has been conjectured (based on simulations) that for connected $d$-regular graphs with $d = \Omega(\log n)$, in $O(dn)$ flips the graph should be transformed into an expander [22]. The recent result of Jacob et al. [17], who give a very different approach and a more involved algorithm (discussed further in the related work section), gives further evidence for this conjecture.

**Our Contribution.**  In this paper we partially bridge the gap between the theoretical results and the experimental results for the flip transform. In particular, we focus on the following natural instantiation of the flip transformation that we call the *random flip*. Our instantiation is only slightly different from the original definition of "flip" (from [22]) in how we randomly select a length-three path. Loosely speaking, we uniformly at random select an edge $(r, s)$ first from the graph, and then select $i$ as a random neighbor of $r$, $j$ as a random neighbor of $s$, 'conditioned on' the four vertices being distinct. In this way we obtain a length-three path $i - r - s - j$.

We show that, starting from any connected $d$-regular connected graph with $d = \Omega(\log n)$, repeatedly applying the random-flip transformation results in a $d$-regular *algebraic expander* in only $O(n^2 d^2 \sqrt{\log n})$ steps, with high probability. A $d$-regular graph is known as an algebraic expander if the nonzero eigenvalues of its Laplacian matrix is at least $\Omega(d)$. It implies that the graph is also a *combinatorial expander*: for every vertex cut that partitions the graph into $t$ and $n - t$ vertices (where $t \leq n/2$), there must be at least $\Omega(dt)$ edges across this cut.

Our analysis differs significantly from all the previous results in this area. Instead of analyzing the Markov chain defined by the random flip, or reducing random flip to simple switch, we study how the eigenvalues of the *graph itself* evolve over time. We show that after $O(n^2 d^2 \log n)$ steps, all the nonzero eigenvalues of the Laplacian of the obtained graph are $\Omega(d)$, with high probability.

A natural attempt to prove such a statement —say, the easier combinatorial expansion property— is to show that, in expectation, the number of edges across each 'small' cuts increases after one step of the random flip. Unfortunately, this is false. There are examples where this number first decreases and then increases.

To better understand how the random flip affects the adjacency matrix (or equivalently, the Laplacian matrix) of a graph $G$, we compute a recursive equation that captures, in expectation, the evolution of the graph from one step to another. Surprisingly, we observe that the expected adjacency matrix of $G$, after one random-flip transformation, is a linear combination of the adjacency matrix of $G$ and its square. This fact suggests that the connectivity of $G$ 'improves' in expectation after each random- flip transformation. However, because this is not a linear recursive formula (as the squared adjacency matrix has appeared), the linearity of expectation does not apply. In other words, this observation does not imply that, even in expectation and even with constant probability, the transformed graph is an expander.

To overcome this difficulty we consider, for analysis purpose only, a potential function based on the matrix exponential of the Laplacian matrix of the current graph $G$. This potential function allows us to keep track of, in a smooth manner, the evolution of the second smallest eigenvalue of the Laplacian of $G$. By applying a recent result on higher-order Cheeger inequality by Kwok *et al.* [19], we are able to prove that our potential evolves nicely and essentially reduces by a factor of $1 - O(d^{-2} n^{-2})$. This is how we show that by repeatedly applying random flips roughly $O(d^2 n^2)$ times any connected $d$-regular graph we can obtain an (algebraic) expander.

Interestingly, to the best of our knowledge, our result is the first direct application of higher-order Cheeger inequality to distributed networks or evolving graphs. We expect to see in the future more results in line with this work.

Finally, we also note that our proof technique generalizes to analyze a natural instantiation of the simple-switch transformation that we call the *random switch*. We show that, starting from any connected $d$-regular connected graph with $d = \Omega(\log n)$, by repeatedly applying the random-switch transformation, it is possible to obtain a $d$-regular *algebraic expander* in only $O(nd)$ steps.

**Related work.** Maintaining good connectivity and a small diameter is a very important task in P2P networks [9]. For this reason, there has been a lot of attention towards building dynamic well-connected networks. In [27], Pandurangan, Raghavan and Upfal proposed a simple scheme to build a regular P2P network with a small diameter. Subsequently Law and Siu [20] obtained an algorithm to preserve the random structure of the graph using local operations in the case when nodes can be added or removed from a P2P network. Our work is similar in spirit to theirs, but our protocol converges to an expander network independently of the properties of the input network.

Random graphs constructed by centralized algorithms or non-local algorithms are also used for building communication networks [10, 25, 32, 33]. In practice, for example for Bittorrent [5] or Gnutella [14], quasi-local hybrid techniques are often preferred.

The recent breakthrough result by Jacob et al. [17] showed how to add edges in a local way and convert any graph into one that has a constant degree expander as a subgraph, in time $O(nd \log n)$. While the result is very powerful, their transformation algorithm is quite involved. Thus, the use of simple procedures like the random flip (which has other advantages such as parallelizability) is still very desirable. Our conclusion is also slightly stronger: we prove that the graph obtained is an algebraic expander — a spectral guarantee as opposed to showing that there is a well-expanding subgraph.

When nodes are added or deleted from a P2P system, some authors also study preserving the connectivity properties of the network. For instance, Pandurangan and Trehan [29] maintains expansion of any graph in a dynamic environment with node deletions and additions. Pandurangan et al. [28] maintains deterministic expanders (p-cycle) starting from some topologies, which have guaranteed expansion in a dynamic environment with node additions and deletions.

Our work also bears resemblance to some of the literature on constructing expanders. An example of this is the Zig-Zag product [30], where an expander is constructed by repeatedly combining two graphs in a tensor-product manner. Another example of this is the so-called "2-lift" [4, 23], an important tool towards explicit constructing bipartite expanders. Incremental algorithms have also been studied towards how to add edges one by one to form an expander (or more generally, a sparsifier) [1, 3]. Of course, random graphs are naturally expanders (see for instance [6]). However, all the cited constructions are non-local.

## 2 Notations

We make use of a number of standard objects in spectral graph theory and linear algebra. All our graphs are undirected, simple and unweighted. They all lie over the same vertex set $V = [n]$. The set $N_G(r)$ for $r \in V$ denotes the set of vertices that are adjacent to $r$ in $G$, while the graph $S_G(r)$ indicates the subgraph of $G$ formed by all edges incident to $r$, i.e. the star graph rooted at $r$ over $N_G(r)$. The graph $K_V$ indicates the complete graph over $V$.

For a graph $G = (V, E)$, we denote by $L(G)$ and $A(G)$ respectively the (unnormalized) Laplacian and the adjacency matrix of $G$. For vertices $u, v \in V$, we write $\delta_{uv} \stackrel{\text{def}}{=} \mathbf{e}_u - \mathbf{e}_v$, where $\mathbf{e}_i$ is the $i$-th standard unit vector, and let $L_{uv} \stackrel{\text{def}}{=} \delta_{uv}\delta_{uv}^T$ be the Laplacian of the graph consisting only of edge $\{u, v\}$. It is clear from this definition that $L(G) = \sum_{e \in E} L_e$.

We denote by $\hat{I} \stackrel{\text{def}}{=} I - \frac{\mathbf{1}\mathbf{1}^T}{n}$, the projection matrix onto the $(n-1)$-dimensional subspace orthogonal to the all-one vector $\mathbf{1}$, and we denote by $\text{span}(\mathbf{1})^\perp$ this subspace. In other words, $\hat{I}$ is the identity matrix in the subspace $\text{span}(\mathbf{1})^\perp$. Notice that $\hat{I} = \frac{1}{n}L(K_V)$, and $\hat{I}L(G) = L(G)\hat{I} = L(G)$ for every graph Laplacian matrix $L(G)$. We also denote by $\hat{\text{tr}}(M)$ and $\hat{e}^M$ the trace and exponential operators in $\text{span}(\mathbf{1})^\perp$.[1]

Finally, we write $A \succeq 0$ if the square matrix $A$ is positive semidefinite (PSD), and $A \succeq B$ if $A - B \succeq 0$.

## 3 Our Model

### 3.1 The Neighbor-Exchange Primitive

Both the random-switch and the random-flip transformations we use in this paper shall be defined on the same *neighbor-exchange primitive* applied to a graph $G = (V, E)$ and a pair $(r, s)$ of distinct vertices in $V$. The execution of this primitive leads $r$ and $s$ to each select and exchange a random neighbor. We call this basic primitive NExchange($G, r, s$). We will see that for the random-switch transformation, $(r, s)$ is sampled from the set of all pairs of distinct vertices, e.g., the edge set of $L(K_V)$. For the random-flip transformation, $(r, s)$ is sampled from the edge of the current graph $G$.

---

[1] As an example, for a graph Laplacian matrix $L(G)$ whose eigenvalues are $0 = \lambda_1 \leq \lambda_2 \leq \cdots \leq \lambda_n$, we have $\hat{\text{tr}}(L(G)) = \lambda_2 + \cdots + \lambda_n$ and $\hat{\text{tr}}(e^{L(G)}) = e^{\lambda_2} + \cdots + e^{\lambda_n}$.

An important challenge in defining NExchange concerns how to deal with the possible creation of multiple edges. In our version of NExchange, we explicitly do not allow multiple edges, as this allows us to preserve the simplicity and regularity of the graph. To implement this feature, we need to introduce a basic fact about regular graphs. For two vertices, $u, v \in V$, define the *disjoint neighborhood* of $u$ with respect to $v$ in $G$ as follows:

$$\Delta_{G,v}(u) \stackrel{\text{def}}{=} N_G(u) \setminus (N_G(v) \cup \{v\}) \ .$$

That is, $\Delta_{G,v}(u)$ is the set of neighbors of $u$ excluding $v$ and $v$'s neighbors. The following fact is a straightforward consequence of the regularity of a graph.

**Fact 3.1.** *For a $d$-regular graph $G = (V, E)$ and vertices $u, v \in V$:*

$$|\Delta_{G,v}(u)| = |\Delta_{G,u}(v)| \stackrel{\text{def}}{=} D_G(u,v) \in \{0, 1, \ldots, d-1\}.$$

We are now ready to formally define the NExchange primitive. On a graph $G$ and a pair of vertices $(r, s)$, the NExchange primitive is described in Algorithm 1. Note that it is a protocol between $r$ and $s$ which can be executed *locally*.

---

**procedure** NExchange($G, r, s$)

**begin**

1.    $r$ samples a vertex $i$ uniformly over its neighborhood $N_G(r)$, and sends it across to $s$.
2.    **if** $i$ is in $N_G(s)$ or $i = s$ **then** the protocol aborts and return the original graph $G$;
3.    **else**
4.       **repeat**
5.          $s$ samples a neighbor $j$ uniformly from $N_G(s)$ and sends it across to $r$.
6.       **until** $j \notin N_G(r)$ and $j \neq r$;
   **end**
7.    $r$ and $s$ replace $i$ and $j$ in their neighborhoods with $j$ and $i$ respectively; we return this new graph $G' = (V, E')$.

**end**

**Algorithm 1:** The NExchange primitive

---

Informally, the NExchange($G, r, s$) works as follows: $r$ first initiates the neighbor exchange process. If the procedure does not terminate in Line 2, then $r$ and $s$ *commit* to making a swap, since the conditions $i \notin N_G(s)$ and $i \neq s$ guarantee that $r$ and $s$ must have at least one non-common neighbor. Furthermore, we observe that conditioned on this, $j$ is picked uniformly from $\Delta_{G,r}(s)$ because every vertex is equally likely to be picked.

We proceed to state a number of basic properties about the NExchange primitive.

**Proposition 3.2.** *Let $G = (V, E)$ be a $d$-regular graph over $n$ vertices. The procedure $G' =$ NExchange$(G, r, s)$ obeys the following properties:*
1. *The output graph $G'$ is simple, connected, and $d$-regular.*
2. *With probability $1 - \frac{D_G(r,s)}{d}$, the procedure aborts and thus does not change the graph.*
3. *With the remaining probability (i.e., $\frac{D_G(r,s)}{d}$), vertices $i$ and $j$ are uniformly and independently distributed over $\Delta_{G,s}(r)$ and $\Delta_{G,r}(s)$ respectively.[2]*

---

[2]In other words, each pair $i \in \Delta_{G,s}(r)$ and $j \in \Delta_{G,r}(s)$ is selected with probability $\frac{1}{d \cdot D_G(r,s)}$.

We are now ready to state the main lemma regarding the NExchange primitive, which gives us explicit bounds on the expected change in the graph Laplacian when NExchange is performed. Its proof requires some standard manipulations of the Laplacian of a graph, and deferred to Appendix A.

**Lemma 3.3.** *Fix a $d$-regular graph $G = (V, E)$ and distinct vertices $r, s \in V$. Let $G'$ be the graph output by* NExchange $(G, r, s)$ *and define $\Delta = L(G') - L(G)$. Then:*

1. $-2\hat{I} \preceq \Delta \preceq 2\hat{I}$ and $\Delta^2 \preceq 4\hat{I}$.

2. $\mathbb{E}[\Delta] = \frac{1}{d} \left( A(G)L_{rs} + L_{rs}A(G) + \mathbb{1}[(r,s) \in E] \cdot 2L_{rs} \right)$.

3. $\mathbb{E}[\Delta^2] \preceq 8L_{rs} + \frac{6}{d} \left( L(S_r) + L(S_s) \right)$.

*Above, $\mathbb{1}[(r,s) \in E]$ is the indicator function that is $1$ if $(r,s)$ is already an edge in $G$.*

## 3.2 Random Switch and Random Flip

Starting from a $d$-regular graph $G^{(0)} = (V, E)$, the random-switch and the random-flip transformations both produce a sequence of $d$-regular graphs $G^{(1)}, G^{(2)}, \ldots, G^{(T)}$ over $V$ by iteratively picking two vertices $r, s \in V$ at random and performing the exchange $G^{(t+1)} = \text{NExchange}(G^{(t)}, r, s)$. The two transformations only differ in how $r$ and $s$ are sampled. At each iteration $t$, we have:

- In the *random-switch* transformation, $r$ and $s$ are chosen uniformly at random among distinct (unordered) pairs of vertices in $V$. Therefore, $\mathbb{E}[L_{rs}] = \frac{1}{\binom{n}{2}} L(K_V)$.

- In the *random-flip* transformation, $(r, s)$ is chosen uniformly at random over the (undirected) edges of $G^{(t)}$. Therefore, $\mathbb{E}[L_{rs}] = \frac{2}{nd} L(G^{(t)})$.

**Basic Bounds on the Laplacian.** The rest of this section is dedicate to proving some basic bounds on the expected behaviors of the random-switch and the random-flip transformations. For this purpose, we write $L^{(t)}$ and $A^{(t)}$ for the Laplacian $L(G^{(t)})$ and the adjacency matrix $A(G^{(t)})$. We also define $\Delta^{(t)} \stackrel{\text{def}}{=} L^{(t+1)} - L^{(t)}$. Using the results of Lemma 3.3, it is straightforward to prove the following lemma on the expectation of $\Delta^{(t)}$ given $G^{(t)}$ for the random-switch transformation, and its proof is deferred to Section B.

**Lemma 3.4.** *Expected behavior of* **Random Switch**

$$\mathbb{E}[\Delta^{(t)}|G^{(t)}] = \frac{4}{n(n-1)} L(K_V) - \frac{4}{nd} L^{(t)} \text{ , and}$$

$$\mathbb{E}[(\Delta^{(t)})^2|G^{(t)}] \preceq \frac{16}{n(n-1)} L(K_V) + \frac{24}{dn} L^{(t)} \text{ .}$$

The first equation established by Lemma 3.4 is particularly useful in understanding the expect behavior of Random Switch: at every iteration it removes a fraction of the current graph and replaces it by a corresponding fraction of the complete graph. Therefore, in expectation, we should arrive at a random $d$-regular expander: that is, $\mathbb{E}[L^{(T)}]$ is very close to $\frac{d}{n-1} L(K_V)$ for $T = \Omega(nd)$ by repeatedly applying . However, it is non-trivial to prove that all the eigenvalues of $L^{(T)}$ are $\Omega(d)$ with descent probability.

One can similarly prove (see Appendix B) the following lemma for the random-flip transformation.

**Lemma 3.5.** *Expected behavior of* **Random Flip**

$$\mathbb{E}[\Delta^{(t)}|G^{(t)}] = \frac{4}{d^2 n} \left( (d+1)L^{(t)} - (L^{(t)})^2 \right) \text{ , and } \quad \mathbb{E}[(\Delta^{(t)})^2|G^{(t)}] \preceq \frac{40}{dn} L^{(t)} \text{ .}$$

An interpretation of the expected behavior of Random Flip can be given as follows. Consider the *squared graph* $G_2^{(t)}$ over $V$, which is formed by considering as edges all paths of length 2 in $G^{(t)}$ excluding self-loops. The Laplacian matrix of $G_2^{(t)}$ is $L(G_2^{(t)}) = d^2 I - A^2$. Then, it is easy to see that, for the random-flip transformation,

$$\mathbb{E}[\Delta^{(t)}|G^{(t)}] = \frac{4}{d^2 n}\left(L(G_2^{(t)}) - (d-1)L(G^{(t)})\right) .$$

In other words, the expected behavior of the Random Flip process is to remove a fraction of the current graph and replace it by a corresponding fraction of its squared graph. This also justifies the convergence of the process to an expander graph, as repeated squaring of a graph yields an expander.

Unfortunately, unlike the random-switch transformation, one cannot repeatedly apply Lemma 3.5 to conclude even $\mathbb{E}[L^{(T)}]$ is close to $\frac{d}{n-1}L(K_V)$ (or equivalently, $\mathbb{E}[L^{(T)}]$ has all its eigenvalues being $\Omega(d)$). It is therefore even harder to conclude that $L^{(T)}$ has all its eigenvalues being $\Omega(d)$ with high probability. This is why the random-flip transformation is particularly challenging to analyze.

## 4 Convergence Analysis for Random Flip

An important component in our analysis is the use of a potential function based on matrix exponentials. At each time $t$, we consider a parameter $\eta_t > 0$ (to be chosen appropriately), and consider the potential:

$$\Phi^{(t)} = \hat{\mathrm{tr}}\left(\hat{e}^{-\eta_t L^{(t)}}\right) .$$

Note that a potential function like this is also known as the matrix moment generating function, and used by the proofs of all the versions of matrix concentration inequalities (see for instance the survey by Tropp [34]).

Like those concentration-inequality proofs, if we manage to prove that the potential $\Phi^{(T)}$ is very small for some $T$, then it implies all the nonzero eigenvalues of $L^{(T)}$ must be large (and therefore $L^{(T)}$ is an expander). However, as discussed at the end of Section 3.2, we do not even know whether the expectation $\mathbb{E}[L^{(T)}]$ has good eigenvalues for large $T$. Therefore, it is irrelevant to apply any matrix concentration bound here.

In this section, we will use a *uniform* value for $\eta_t = \eta$ for all the iterations $t$. The key technical lemma, stated below, implies that for any graph which is not yet an expander, the potential drops in expectation by a considerable amount. This allows us to bound the number of steps needed to obtain an expander.

The proof of this lemma, deferred to Section 4.1, is closely related to the Matrix Multiplicative Weight Updates algorithm [1, 2, 26] where the matrix losses are given by the differences $\Delta^{(t+1)} \stackrel{\text{def}}{=} L^{(t+1)} - L^{(t)}$.

**Lemma 4.1.** *Let $\eta \stackrel{\text{def}}{=} 20 \log n / d$. If $\eta \leq \frac{1}{20}$, then for any d-regular graph $G^{(t)}$, we have*

$$\mathbb{E}[\Phi^{(t+1)} \mid G^{(t)}] \leq \left(1 - O\left(\frac{\sqrt{\log n}}{d^2 n^2}\right)\right)\Phi^{(t)} + O(n^{-3}) .$$

This above lemma easily implies our main result:

**Theorem 4.2.** *If $d \geq 100 \log n$, after $T = \Omega(d^2 n^2 \sqrt{\log n})$ random flips, with probability at least $1 - n^{-2}$, we have that the smallest nonzero eigenvalue of $L^{(T)}$ is at least $\Omega(d)$, or in other words, $G^{(T)}$ is an (algebraic) expander.*

*Proof.* Recall that $\Phi^{(0)} \leq n - 1 < n$ by definition. Therefore, repeated applications of Lemma 4.1 imply that after $T = \Omega(d^2 n^2 \sqrt{\log n})$ random flips, we have that $\mathbb{E}[\Phi^{(T)}] \leq O(n^{-3})$. Thus by Markov's inequality, we have that with probability at least $1 - 1/n^2$, $\Phi^{(T)} \leq O(1/n)$. In particular, the smallest non-zero eigenvalue of $G^{(T)}$ satisfies

$$\lambda_{\min}(L^{(T)}) = \frac{-\log(e^{-\eta \lambda_{\min}(L^{(T)})})}{\eta} \geq -\frac{\log(\Phi^{(T)})}{\eta} \geq \Omega(d) .$$

□

We now get to the core of the analysis, which is to prove the Lemma 4.1.

## 4.1 Proof of Lemma 4.1

We start by observing that

$$\Phi^{(t+1)} = \hat{\mathrm{tr}}\big(\hat{e}^{-\eta L^{(t+1)}}\big) = \hat{\mathrm{tr}}\big(\hat{e}^{-\eta(L^{(t)}+\Delta^{(t)})}\big) \leq \hat{\mathrm{tr}}\big(\hat{e}^{-\eta L^{(t)}}\hat{e}^{-\eta\Delta^{(t)}}\big) \;, \tag{4.1}$$

where the last inequality is due to the Golden-Thompson inequality that says $\mathrm{tr}(e^{A+B}) \preceq \mathrm{tr}(e^A e^B)$ for symmetric matrices $A$ and $B$. Recall that for any flip, we have that $-2\hat{I} \preceq \Delta^{(t)} \preceq 2\hat{I}$ according to Lemma 3.3. This in turn implies that for any $\eta \leq 1/2$, we have[3]

$$\hat{e}^{-\eta\Delta^{(t)}} \preceq \hat{I} - \eta\Delta^{(t)} + \eta^2(\Delta^{(t)})^2 \;.$$

Plugging this into (4.1), and using the fact that for PSD matrices $A, B, C$ with $B \preceq C$, we have $\mathrm{tr}(AB) \leq \mathrm{tr}(AC)$, we get

$$\Phi^{(t+1)} \leq \hat{\mathrm{tr}}\left(\hat{e}^{-\eta L^{(t)}}\left(I - \eta\Delta^{(t)} + \eta^2\left(\Delta^{(t)}\right)^2\right)\right) \;.$$

Next, using Lemma 3.5, we can bound

$$\mathbb{E}[\Phi^{(t+1)}|G^{(t)}] \leq \Phi^{(t)} - \frac{4\eta}{d^2 n} \cdot \hat{\mathrm{tr}}\left(\hat{e}^{-\eta L^{(t)}}\left((d+1)L^{(t)} - \left(L^{(t)}\right)^2 - 10\eta d L^{(t)}\right)\right)$$

$$= \Phi^{(t)} - \frac{4\eta}{d^2 n} \cdot \hat{\mathrm{tr}}\left(\hat{e}^{-\eta L^{(t)}}\left((d+1-10\eta d)L^{(t)} - \left(L^{(t)}\right)^2\right)\right)$$

Since we have chosen $\eta \leq 1/20$, we have

$$\mathbb{E}[\Phi^{(t+1)}|G^{(t)}] \leq \Phi^{(t)} - \frac{4\eta}{d^2 n} \cdot \hat{\mathrm{tr}}\left(\hat{e}^{-\eta L^{(t)}}\left(L^{(t)}\left(\frac{d}{2}\hat{I} - L^{(t)}\right)\right)\right) \;. \tag{4.2}$$

Let us analyze the trace term above. Since all the matrices have a common diagonalization, we can write out the trace in terms of the eigenvalues of $L^{(t)}$. We denote them by $0 = \lambda_1 < \lambda_2 \leq \cdots \leq \lambda_n \leq 2d$, and recall that $\sum_i \lambda_i = nd$. The trace term above is then equal to

$$\sum_{2 \leq i \leq n} e^{-\eta\lambda_i}\lambda_i(d/2 - \lambda_i). \tag{4.3}$$

Note that the factor $(d/2 - \lambda_i)$ is not always positive, so some of these summands can contribute negatively to the drop in the potential. This is the place we use our choice of $\eta$ to conclude that the exponential factor 'kills' the negative contribution.

Formally, consider the following two cases: (1) all the nonzero eigenvalues of $G^{(t)}$ are already greater than or equal to $d/4$, or (2) at least one of them is smaller than $d/4$.

In the former case, we have that the quantity (4.3) is at least $-ne^{-\eta d/4} \cdot 3d^2 \geq -3n^{-2}$, owing to our choice of $\eta = 20\log n/d$. In this case, we conclude that $\mathbb{E}[\Phi^{(t+1)}|G^{(t)}] \leq \Phi^{(t)} + \frac{6\eta}{d^2 n^3} \leq \Phi^{(t)} + O(n^{-3})$. Furthermore in this case for our choice of $\eta$, we have $\Phi^{(t)} \in O(n^{-3})$ and so $\mathbb{E}[\Phi^{(t+1)}|G^{(t)}] \leq O(n^{-3})$.

The rest of this section is to deal with the latter case, that is, when at least one of the nonzero eigenvalues (and in particular, $\lambda_2$) is smaller than $d/4$. Let $m \geq 3$ be the smallest index for which $\lambda_m \geq d/4$. Using the choice of $\eta = 20\log n/d$ again, we have that for the large indices

$$\sum_{i \geq m} e^{-\eta\lambda_i}\lambda_i(d/2 - \lambda_i) \geq -n \cdot e^{-d\eta/4} \cdot 3d^2 \geq 3n^{-2} \;. \tag{4.4}$$

---

[3]This follows from the inequality for real numbers: $e^{-z} \leq 1 - z + z^2$ for $|z| \leq 1$.

For the smaller indices, we have

$$\sum_{2 \leq i < m} e^{-\eta \lambda_i} \lambda_i (d/2 - \lambda_i) \geq \frac{d}{4} \sum_{2 \leq i < m} e^{-\eta \lambda_i} \lambda_i \ . \tag{4.5}$$

We now lower bound on the right hand side of inequality (4.5) relative to the current potential $\Phi^{(t)} = \sum_{2 \leq i \leq n} e^{-\eta \lambda_i}$. More precisely, we make the following observation

**Claim 4.3.** $\frac{\sum_{2 \leq i < m} e^{-\eta \lambda_i} \lambda_i}{\sum_{2 \leq i \leq n} e^{-\eta \lambda_i}} \geq \Omega\left(\frac{1}{n\sqrt{\log n}}\right).$

Before proving the claim, let us note that combining it with (4.4), (4.5), we obtain

$$\sum_{2 \leq i \leq n} e^{-\eta \lambda_i} \lambda_i (d/2 - \lambda_i) \geq \Omega\left(\frac{d}{n\sqrt{\log n}}\right) \cdot \Phi^{(t)} - O\left(\frac{1}{n^2}\right) \ .$$

Plugging this back to (4.2) and using our choice of $\eta$, we arrive at

$$\mathbb{E}[\Phi^{(t+1)}|G^{(t)}] \leq \left(1 - O\left(\frac{\sqrt{\log n}}{d^2 n^2}\right)\right)\Phi^{(t)} + O(n^{-3}) \ .$$

This finishes the proof of Lemma 4.1. Thus, it suffices to show the claim above.

## 4.2 Proof of Claim 4.3

From our argument above, we have that $\lambda_2 \leq d/4$. However, we can assume something much stronger. If $\lambda_2 \geq 1/\eta = d/(20 \log n)$, then the ratio on the left hand side in Claim 4.3 is at least $\lambda_2/n \geq 1/n$, and there is nothing to prove. Therefore, we can assume without loss of generality that $\lambda_2 < \frac{1}{\eta} = \frac{d}{20 \log n}$.

In what follows, let $k$ denote the smallest index for which $\lambda_k \geq 1/\eta$, and the above discussion implies $k \geq 3$.

The rest of the proof is divided into two parts. We first show that Claim 4.3 is implied by the following inequality (4.6), and then prove (4.6).

$$\frac{\sum_{2 \leq i < k} e^{-\eta \lambda_i} \lambda_i}{\sum_{2 \leq i < k} e^{-\eta \lambda_i}} \geq \Omega\left(\frac{1}{n\sqrt{\log n}}\right) \ . \tag{4.6}$$

**Proving (4.6) $\implies$ Claim 4.3.** This is done by massaging the numerator and the denominator of the left hand side of Claim 4.3.

We first notice that $e^{-\eta \lambda_2} \geq 1/e$ since $\lambda_2 < 1/\eta$. Also, by the definition of $m$, we know $\sum_{i \geq m} e^{-\eta \lambda_i} \leq n e^{-\eta d/4} \leq 1/n^4$. Therefore, we have $\sum_{2 \leq i < m} e^{-\eta \lambda_i} \geq (1/2)\left(\sum_{2 \leq i \leq n} e^{-\eta \lambda_i}\right)$. For this reason, to prove Claim 4.3, it suffices to show that

$$\frac{\sum_{2 \leq i < m} e^{-\eta \lambda_i} \lambda_i}{\sum_{2 \leq i < m} e^{-\eta \lambda_i}} \geq \Omega\left(\frac{1}{n\sqrt{\log n}}\right) \ . \tag{4.7}$$

Next, recall that $k$ is the smallest integer such that $\lambda_k \geq 1/\eta$, and $m \geq k \geq 3$. We now want to prove that (4.6) implies (4.7). If $k = m$ then (4.7) is the same as (4.6). Otherwise (i.e., $k < m$), since $\lambda_i$ are in increasing order, we have

$$\frac{\sum_{k \leq i < m} e^{-\eta \lambda_i} \lambda_i}{\sum_{k \leq i < m} e^{-\eta \lambda_i}} \geq \lambda_k \geq \Omega\left(\frac{d}{\log n}\right) \geq \Omega\left(\frac{1}{n\sqrt{\log n}}\right). \tag{4.8}$$

By simple averaging, if $x_1/y_1$ and $x_2/y_2$ are both $\Omega(1/n\sqrt{\log n})$, then so is the ratio $(x_1 + x_2)/(y_1 + y_2)$. Therefore, (4.6), together with (4.8) above, implies (4.7) and thus also Claim 4.3.

Thus, to prove Claim 4.3, we only need to show that (4.6) holds.

**Proving Inequality (4.6).** We view the left hand side of (4.6) as a weighted average over the $\lambda_i$'s for $2 \leq i < k$, where each $\lambda_i$ has weight $e^{-\eta \lambda_i}$. However, since $\lambda_i \leq \lambda_2 < 1/\eta$, we have that $e^{-\eta \lambda_i} \in [e^{-1}, 1]$ is a constant for every $2 \leq i < k$. For this reason, to prove (4.6) it suffices to show that the unweighted average over $\lambda_i$'s for $2 \leq i < k$ is at least $\Omega\left(\frac{1}{n\sqrt{\log n}}\right)$: that is,

$$\frac{\sum_{2 \leq i < k} \lambda_i}{k - 2} \geq \Omega\left(\frac{1}{n\sqrt{\log n}}\right). \tag{4.9}$$

This is simply a statement about the average of the small non-zero eigenvalues of the Laplacian, and is of independent interest. We prove it by using some of the recent beautiful works on higher-order Cheeger inequality of graph Laplacians. Specifically, the next theorem is due to Kwok *et al.* [19], restated below for our convenience:

**Theorem 4.4** (Corollary 1.3 in [19], restated)**.** *For any $d$-regular undirected graph $G = (V, E)$, and $q > p \geq 2$, we have*

$$\phi_p(G) \leq O(qp^6) \frac{\lambda_p/d}{\sqrt{\lambda_q/d}}. \tag{4.10}$$

*In addition, if $p \geq 3$, we have*

$$\phi_{\lceil p/2 \rceil}(G) \leq O\left(\frac{q \log^2 p}{p}\right) \cdot \frac{\lambda_p/d}{\sqrt{\lambda_q/d}}. \tag{4.11}$$

*Above, $0 = \lambda_1 \leq \cdots \leq \lambda_n \leq 2d$ are the eigenvalues of the (unnormalized) Laplacian of $G$, and for integer $t \geq 2$, $\phi_t(G)$ is the minimum over all possible $t$-partitions $(V_1, V_2, \ldots, V_t)$ of the vertex set $V$, of the quantity*

$$\max_{j \in [t]} \frac{E(V_t, V \setminus V_t)}{d|V_t|}.$$

Indeed, if $k < 6$, then by plugging in $q = k$ and $p = k-1$ to (4.10), we have that $\lambda_{k-1}/d \geq \Omega(\phi_{k-1}(G)) \cdot \sqrt{\lambda_k/d}$. Combining this with the fact that $\phi_{k-1}(G) \geq 1/dn$ for any connected graph as well as the choice that $\lambda_k \geq \Omega(\frac{d}{\log n})$, we have that $\lambda_{k-1} \geq \Omega\left(\frac{1}{n\sqrt{\log n}}\right)$. This, together with the assumption that $k$ is a constant, yields (4.9).

If $k \geq 6$, we plug in $p = \lceil k/2 \rceil$ and $q = k$ in (4.11). Together with the choice that $\lambda_k \geq \Omega(\frac{d}{\log n})$, we have

$$\lambda_{\lceil k/2 \rceil}/d \geq \Omega\left(\phi_{\lceil \lceil k/2 \rceil/2 \rceil}(G)/(\log^2 k \log^{0.5} n)\right). \tag{4.12}$$

Now, how small can $\phi_t$ be for any $t \geq 2$? We use again the connectivity of the graph in each iteration. For any partition of $V$ into $t$ pieces, we have that $|E(V_j, V \setminus V_j)| \geq 1$, and for at least one piece, we must have $|V_j| \leq n/t$, which gives that $\phi_t \geq t/nd$. Plugging this into (4.12), we have that

$$\lambda_{\lceil k/2 \rceil} \geq \Omega\left(\frac{k/\log^2 k}{n\sqrt{\log n}}\right) \geq \Omega\left(\frac{1}{n\sqrt{\log n}}\right).$$

Since the average of $\lambda_2, \lambda_3, \ldots, \lambda_{k-1}$ is at least $\Omega(\lambda_{\lceil k/2 \rceil})$, we conclude that (4.9) is proven. This finishes the proof of Claim 4.3, and thus the proof of our main result. □

# 5 Convergence Analysis for Random Switch

Our proof follows the same general outline as the one for the random-flip. We will maintain an exponential potential function, but in this case, it turns out to be more convenient to use *non-uniform* values of $\eta_t$ in our potential function. That is, at time step $t$, we define

$$\Phi^{(t)} = \hat{\text{tr}}\left(\hat{e}^{-\eta_t L^{(t)}}\right) .$$

The choice of the $\eta_t$'s is inspired by the expression for $\mathbb{E}[L^{(t+1)}|G^{(t)}]$ derived in Lemma 3.4, i.e.,

$$\mathbb{E}[L^{(t+1)}|G^{(t)}] = \left(1 - \frac{4}{nd}\right)L^{(t)} + \frac{4}{n(n-1)}L(K_V) .$$

The intuition is as follows. As described in Section 2, the matrix $L(K_V)$ above behaves the same as $n$ times the identity matrix $\hat{I}$ (in the space orthogonal to $\mathbf{1}$). Thus, if $L^{(t+1)}$ were always equal to its expectation conditioned on $G^{(t)}$, we could have set $\eta_t = (1 - 4/nd)\eta_{t+1}$ and have $\hat{e}^{-\eta_{t+1}L^{(t+1)}} = \hat{e}^{-\eta_t L^{(t)} - \frac{4\eta_{t+1}}{n-1}\hat{I}} = \hat{e}^{-\eta_t L^{(t)}} \cdot e^{-4\eta_{t+1}/(n-1)}$. This implies that the potential decreases by a factor roughly $e^{-4\eta_{t+1}/(n-1)} \approx \left(1 - \frac{4\eta_{t+1}}{n-1}\right)$.

To turn this intuition into a formal proof, we choose a slightly different ratio between $\eta_t$ and $\eta_{t+1}$. We pick them so as to satisfy:

$$\eta_t = \left(1 - \frac{8}{d(n-1)}\right)\eta_{t+1} \text{ for all } t = 0, 1, \ldots, T-1, \quad \eta_T = 1/6, \quad \text{and } T = \left\lceil \frac{dn}{8} \right\rceil .$$

Note that this choice implies $\eta_t \le \frac{1}{6}$ for all the values $t \le T$. Our main technical lemma is as follows.

**Lemma 5.1.** *For the aforementioned choices of $\eta_t$'s, we have that for every $t = 0, 1, \ldots, T-1$,*

$$\mathbb{E}[\Phi^{(t+1)}|G^{(t)}] = \mathbb{E}\left[\hat{\text{tr}}\left(\hat{e}^{-\eta_{t+1}(L^{(t)}+\Delta^{(t)})}\right)|G^{(t)}\right] \le \Phi^{(t)} \cdot e^{-\eta_{t+1}\frac{2}{n-1}} .$$

Before proving Lemma 5.1, let us point out how it implies our main theorem for the random-switch transformation. The lemma implies, by multiplying it out for $t = 0, 1, \ldots, T-1$,

$$\mathbb{E}\left[\hat{\text{tr}}\left(\hat{e}^{-\eta_T L^{(T)}}\right)\right] \le \hat{\text{tr}}\left(\hat{e}^{-\eta_0 L^{(0)}}\right) \cdot e^{-\frac{2}{n-1}\sum_{t=1}^{T}\eta_t} .$$

As a result, we obtain

$$\mathbb{E}\left[e^{-\eta_T \lambda_{\min}(L^{(T)})}\right] \le \mathbb{E}\left[\hat{\text{tr}}\left(\hat{e}^{-\eta_T L^{(T)}}\right)\right] \le \hat{\text{tr}}\left(\hat{e}^{-\eta_0 L^{(0)}}\right) \cdot e^{-\frac{2}{n-1}\sum_{t=1}^{T}\eta_t} \le n \cdot e^{-\frac{2}{n-1}\sum_{t=1}^{T}\eta_t}$$

$$\le n \cdot e^{-\frac{2}{n-1}\eta_T \frac{1-(1-8/d(n-1))^T}{1-(1-8/d(n-1))}} = n \cdot e^{-\frac{2}{n-1}\frac{1}{6}(1-1/e)\frac{d(n-1)}{8}} = n \cdot e^{-\frac{d}{24}(1-1/e)} .$$

Using Markov's inequality, we conclude that with probability at least $1 - \delta$, it satisfies $e^{-\eta_T \lambda_{\min}(L^{(T)})} \le \frac{n}{\delta} \cdot e^{-\frac{d}{24}(1-1/e)}$, or equivalently,

$$\lambda_{\min}(L^{(T)}) \ge \frac{6d}{24}(1 - 1/e) - 6\log(n/\delta) > \frac{d}{8} - 6\log(n/\delta) .$$

In sum, we have the following theorem:

**Theorem 5.2.** *After $T = \lceil \frac{dn}{8} \rceil$ random switches, with probability at least $1 - \delta$, we have $\lambda_{\min}(L^{(T)}) \ge \frac{d}{8} - 6\log(n/\delta)$.*

The theorem implies that, in particular, if $d = \Omega(\log n)$ then $G^{(T)}$ is an (algebraic) expander with probability at least $1 - 1/\text{poly}(n)$.

## 5.1 Proof of Lemma 5.1

The key ingredient in our proof is Lieb's theorem, which states that for any symmetric matrix $A$, the function $f(B) = \mathrm{tr}(e^{A+\log B})$ is concave over the positive cone $B \succeq 0$. Introduced in the context of quantum mechanisms [21], this concavity theorem has recently proven useful in showing matrix concentration bounds [34] and in the analysis of some matrix variants of the multiplicative weight update method [1].

We apply Lieb's theorem as follows. Recall that the expectation of any concave function is no greater than the function applied on its expectation. Therefore, conditioned on $G^{(t)}$ so $\Delta^{(t)}$ is the only random variable, we have

$$\mathbb{E}\big[\hat{\mathrm{tr}}\big(\hat{e}^{-\eta_{t+1}(L^{(t)}+\Delta^{(t)})}\big)\big|G^{(t)}\big] = \mathbb{E}\big[\hat{\mathrm{tr}}\big(\hat{e}^{-\eta_{t+1}L^{(t)}+\log e^{-\eta_{t+1}\Delta^{(t)}}}\big)\big|G^{(t)}\big]$$
$$\leq \hat{\mathrm{tr}}\big(\hat{e}^{-\eta_{t+1}L^{(t)}+\log \mathbb{E}_t[e^{-\eta_{t+1}\Delta^{(t)}}|G^{(t)}]}\big) \tag{5.1}$$

Next, since $-2\hat{I} \preceq \Delta^{(t)} \preceq 2\hat{I}$ according to Lemma 3.3 and $\eta_{t+1} \leq \frac{1}{6}$, we have[4]

$$\mathbb{E}[\hat{e}^{-\eta_{t+1}\Delta^{(t)}}|G^{(t)}] \preceq \mathbb{E}\big[\hat{I} - \eta_{t+1}\Delta^{(t)} + \frac{3}{4}\cdot\eta_{t+1}^2(\Delta^{(t)})^2\big|G^{(t)}\big]$$
$$\preceq \hat{I} - \eta_{t+1}\Big(\frac{4}{n-1}\hat{I} - \frac{4}{d(n-1)}L^{(t)} - \frac{3\eta_{t+1}}{4}\Big(\frac{16}{n-1}\hat{I} + \frac{24}{d(n-1)}L^{(t)}\Big)\Big)$$
$$\preceq \hat{I} - \eta_{t+1}\frac{2}{n-1}\hat{I} + \eta_{t+1}\frac{8}{d(n-1)}L^{(t)} \ .$$

Above, the second inequality is due to Lemma 3.4 and the fact that $L(K_V) = n\hat{I}$, and the third inequality uses our choice that $\eta_{t+1} \leq \frac{1}{6}$. Next, due to the operator monotonicity of the log function, we have

$$\log\mathbb{E}[e^{-\eta_{t+1}\Delta^{(t)}}|G^{(t)}] \preceq \log\big(\hat{I} - \eta_{t+1}\frac{2}{n-1}\hat{I} + \eta_{t+1}\frac{8}{d(n-1)}L^{(t)}\big) \preceq -\eta_{t+1}\frac{2}{n-1}\hat{I} + \eta_{t+1}\frac{8}{d(n-1)}L^{(t)} \tag{5.2}$$

Combining (5.1) and (5.2), as well as the monotonicity of $A \mapsto \mathrm{tr}(e^A)$ over symmetric matrices $A$, we conclude that

$$\mathbb{E}\big[\hat{\mathrm{tr}}\big(\hat{e}^{-\eta_{t+1}(L^{(t)}+\Delta^{(t)})}\big)\big|G^{(t)}\big] \leq \hat{\mathrm{tr}}\big(\hat{e}^{-\eta_{t+1}L^{(t)}-\eta_{t+1}\frac{2}{n-1}\hat{I}+\eta_{t+1}\frac{8}{d(n-1)}L^{(t)}}\big)$$
$$= \hat{\mathrm{tr}}\big(\hat{e}^{-\eta_{t+1}(1-\frac{8}{d(n-1)})L^{(t)}-\eta_{t+1}\frac{2}{n-1}\hat{I}}\big)$$
$$= \hat{\mathrm{tr}}\big(\hat{e}^{-\eta_t L^{(t)}}\big) \cdot e^{-\eta_{t+1}\frac{2}{n-1}} \ . \qquad \square$$

## Acknowledgments

We would to thank Nima Ahmadipouranari, Shayan Oveis Gharan, Bernhard Haeupler, Jelani Nelson and Prasad Tetali for the useful discussions. This material is based upon work partly supported by the National Science Foundation under Grant CCF-1319460.

---

[4]This follows from the inequality for real numbers: $e^{-z} \leq 1 - z + (3/4)z^2$ for $|z| \leq 1/3$.

# APPENDIX

## A Missing Proof of Lemma 3.3

**Lemma 3.3.** *Fix a $d$-regular graph $G = (V, E)$ and distinct vertices $r, s \in V$. Let $G'$ be the graph output by* NEXCHANGE $(G, r, s)$ *and define $\Delta = L(G') - L(G)$. Then:*

1. $-2\hat{I} \preceq \Delta \preceq 2\hat{I}$ *and* $\Delta^2 \preceq 4\hat{I}$.

2. $\mathbb{E}[\Delta] = \frac{1}{d} \left( A(G) L_{rs} + L_{rs} A(G) + \mathbb{1}[(r,s) \in E] \cdot 2 L_{rs} \right)$.

3. $\mathbb{E}[\Delta^2] \preceq 8 L_{rs} + \frac{6}{d} \left( L(S_r) + L(S_s) \right)$.

*Above, $\mathbb{1}[(r,s) \in E]$ is the indicator function that is 1 if $(r,s)$ is already an edge in $G$.*

1. To prove the first item, notice that when NEXCHANGE does not abort and some change takes place, we have $\Delta = L_{is} + L_{jr} - L_{ir} - L_{sj}$. Then, a calculation reveals the eigenvector decomposition of $\Delta$ as:

$$\Delta = \delta_{is}\delta_{is}^T + \delta_{jr}\delta_{jr}^T - \delta_{ir}\delta_{ir}^T - \delta_{js}\delta_{js}^T = \delta_{rs}\delta_{ij}^T + \delta_{ij}\delta_{rs}^T = \frac{1}{2}\left[(\delta_{rs} + \delta_{ij})(\delta_{rs} + \delta_{ij})^T - (\delta_{rs} - \delta_{ij})(\delta_{rs} - \delta_{ij})^T\right].$$

   As the vectors $(\delta_{rs} + \delta_{ij})$ and $(\delta_{ij} - \delta_{rs})$ are orthogonal, they are eigenvectors. Because both of these vectors have square norm 4, the corresponding eigenvalues are 2 and $-2$. The statement $-2\hat{I} \preceq \Delta \preceq 2\hat{I}$ follows as both eigenvectors are orthogonal to the all-one vector. The statement $\Delta^2 \preceq 4\hat{I}$ hold for the same reason.

2. For the second item, let $C$ be the event {NEXCHANGE does not abort}. Then, the formula $\Delta = \delta_{rs}\delta_{ij}^T + \delta_{ij}\delta_{rs}^T$ lets us compute the following:

$$\mathbb{E}[\Delta] = \Pr[C] \cdot \left( \delta_{rs} \mathbb{E}[\delta_{ij}|C]^T + \mathbb{E}[\delta_{ij}|C] \delta_{rs}^T \right). \tag{A.1}$$

   Hence, it suffices to calculate $\mathbb{E}[\delta_{ij}|C]$. For this purpose, consider the vector $A(G)\delta_{rs} + \mathbb{1}[(r,s) \in E] \cdot \delta_{rs}$. It is easy to verify by hand that

$$\forall i \in V, \quad \left(A(G)\delta_{rs} + \mathbb{1}[(r,s) \in E] \cdot \delta_{rs}\right)_i = \begin{cases} 1 & \text{if } i \in N_G(r) \setminus (N_G(s) \cup \{s\}) = \Delta_{G,s}(r), \\ -1 & \text{if } j \in N_G(s) \setminus (N_G(r) \cup \{r\}) = \Delta_{G,r}(s), \\ 0 & \text{otherwise} \end{cases}$$

   In addition, recall from Proposition 3.2 that, conditioned on event $C$, vertices $i$ and $j$ are selected from $\Delta_{G,s}(r)$ and $\Delta_{G,r}(s)$ each with probability $1/D_G(r,s)$. Therefore, $\mathbb{E}[\delta_{ij}|C] = \frac{1}{D_G(r,s)} \cdot (A(G)\delta_{rs} + \mathbb{1}[(r,s) \in E] \cdot \delta_{rs})$. Plugging this into (A.1) and using the fact that $\Pr[C] = \frac{D_G(r,s)}{d}$ (again from Proposition 3.2), we arrive at the desired equality.

3. Finally, for the third item, we can write:

$$\mathbb{E}[\Delta^2] = \Pr[C] \cdot \mathbb{E}[(\delta_{rs}\delta_{ij}^T + \delta_{ij}\delta_{rs}^T)^2 | C] = \frac{D_G(r,s)}{d} \left( \mathbb{E}[2L_{ij}|C] + 2L_{rs} \right)$$

   where the last equality follows from the fact that $\delta_{ij}^T \delta_{rs} = 0$ because $r, s, i$ and $j$ are all distinct. To upper bound $L_{ij}$, notice that $L_{ij} \preceq 3(L_{rs} + L_{ir} + L_{sj})$ for any four vertices $i, r, s, j$.[5] In addition, conditioned on $C$, $\mathbb{E}[L_{ir}|C] \preceq \frac{1}{D_G(r,s)} L(S_r)$ and $\mathbb{E}[L_{sj}|C] \preceq \frac{1}{D_G(r,s)} L(S_s)$. Together, we arrive at the desired inequality. □

---
[5]This follow from the inequality for real numbers: $|a| + |b| + |c| \leq 3(a^2 + b^2 + c^2)$

# B  Missing Proofs in Section 3.2

**Lemma 3.4.** *Expected behavior of* **Random Switch**

$$\mathbb{E}[\Delta^{(t)}|G^{(t)}] = \frac{4}{n(n-1)}L(K_V) - \frac{4}{nd}L^{(t)} \text{ , and}$$

$$\mathbb{E}[(\Delta^{(t)})^2|G^{(t)}] \preceq \frac{16}{n(n-1)}L(K_V) + \frac{24}{dn}L^{(t)} .$$

*Proof of Lemma 3.4.* We first compute that

$$\mathbb{E}[\Delta^{(t)}|G^{(t)}] \stackrel{①}{=} \frac{1}{d}\Big(A^{(t)}\mathbb{E}[L_{rs}|G^{(t)}] + \mathbb{E}[L_{rs}|G^{(t)}]A^{(t)} + 2\mathbb{E}[\mathbb{1}[(r,s) \in E] \cdot L_{rs}|G^{(t)}]\Big)$$

$$\stackrel{②}{=} \frac{1}{d}\Big(A^{(t)}\frac{L(K_V)}{\binom{n}{2}} + \frac{L(K_V)}{\binom{n}{2}}A^{(t)} + \frac{2}{\binom{n}{2}}L^{(t)}\Big)$$

$$\stackrel{③}{=} \frac{1}{d\binom{n}{2}}\Big((dI - L^{(t)})n\hat{I} + n\hat{I}(dI - L^{(t)}) + 2L^{(t)}\Big)$$

$$= \frac{4}{n-1}\hat{I} - \frac{4}{dn}L^{(t)} \stackrel{④}{=} \frac{4}{n(n-1)}L(K_V) - \frac{4}{nd}L^{(t)} .$$

Above, ① uses Lemma 3.3, ② follows from the definition of the random-switch transformation, and ③ and ④ follow because $L^{(t)} = dI - A^{(t)}$ and $L(K_V) = n\hat{I}$.

Next, we compute that

$$\mathbb{E}[(\Delta^{(t)})^2|G^{(t)}] \stackrel{⑤}{\preceq} 8\mathbb{E}[L_{rs}|G^{(t)}] + \frac{6}{d}\Big(\mathbb{E}[L(S_r) + L(S_s)|G^{(t)}]\Big) \stackrel{⑥}{=} \frac{8}{\binom{n}{2}}L(K_V) + \frac{24}{dn}L^{(t)} .$$

Above, ⑤ uses Lemma 3.3, and ⑥ uses the definition of the random-switch transformation, as well as the fact that if $r$ (resp. $s$) is uniformly distributed over $V$, then the star graph $L(S_r)$ (resp. $L(S_s)$) has an expectation that equals $\frac{2}{n}L^{(t)}$. □

**Lemma 3.5.** *Expected behavior of* **Random Flip**

$$\mathbb{E}[\Delta^{(t)}|G^{(t)}] = \frac{4}{d^2n}\big((d+1)L^{(t)} - (L^{(t)})^2\big) \text{ , and}$$

$$\mathbb{E}[(\Delta^{(t)})^2|G^{(t)}] \preceq \frac{40}{dn}L^{(t)} .$$

*Proof of Lemma 3.5.* We first compute that

$$\mathbb{E}[\Delta^{(t)}|G^{(t)}] \stackrel{①}{=} \frac{1}{d}\Big(A^{(t)}\mathbb{E}[L_{rs}|G^{(t)}] + \mathbb{E}[L_{rs}|G^{(t)}]A^{(t)} + 2\mathbb{E}[L_{rs}|G^{(t)}]\Big)$$

$$\stackrel{②}{=} \frac{1}{d}\Big(A^{(t)}\frac{L^{(t)}}{dn/2} + \frac{L^{(t)}}{dn/2}A^{(t)} + \frac{2}{dn/2}L^{(t)}\Big)$$

$$\stackrel{③}{=} \frac{2}{d^2n}\big((dI - L^{(t)})L^{(t)} + L^{(t)}(dI - L^{(t)}) + 2L^{(t)}\big) = \frac{4(d+1)}{d^2n}L^{(t)} - \frac{4}{d^2n}(L^{(t)})^2 .$$

Above, ① uses Lemma 3.3 and the fact that $\mathbb{1}[(r,s) \in E]$ is always 1 for the random-flip transformation, ② follows from the definition of the random-flip transformation, and ③ follows because $L^{(t)} = dI - A^{(t)}$.

Next, we compute that

$$\mathbb{E}[(\Delta^{(t)})^2|G^{(t)}] \stackrel{④}{\preceq} 8\mathbb{E}[L_{rs}|G^{(t)}] + \frac{6}{d}\Big(\mathbb{E}[L(S_r) + L(S_s)|G^{(t)}]\Big) \stackrel{⑤}{=} \frac{8}{dn/2}L^{(t)} + \frac{24}{dn}L^{(t)} .$$

Above, ④ uses Lemma 3.3, and ⑤ uses the definition of the random-flip transformation, as well as the fact that if $r$ (resp. $s$) is uniformly distributed over $V$, then the star graph $L(S_r)$ (resp. $L(S_s)$) has an expectation that equals $\frac{2}{n}L^{(t)}$. □